\documentclass[trackchanges]{aastex701}

\usepackage{overpic}
\usepackage{multirow} 
\usepackage{booktabs}

\usepackage{hyperref}
\received{\today}
\revised{\today}
\accepted{\today}
\begin{document}

\title{A Tentative Line-like GeV Excess in $Fermi$ Blazar 4FGL J1754.2$+$3212:\\ Implications for Jet Physics and Beyond}

\author[orcid=0000-0002-9071-5469,gname='Shi-Ju',sname='Kang']{Shi-Ju Kang}
\affiliation{School of Physics and Electrical Engineering, Liupanshui Normal University, Liupanshui, Guizhou, 553004, People's Republic of China}
\email[show]{kangshiju@alumni.hust.edu.cn} 
\email[]{kangshiju@lpssy.edu.cn}

\author[orcid=0009-0008-5422-7485,gname='Yue',sname='Yin']{Yue Yin}
\affiliation{School of Physics and Electrical Engineering, Liupanshui Normal University, Liupanshui, Guizhou, 553004, People's Republic of China}
\email[]{yueyin@lpssy.edu.cn}

\author[orcid=0000-0003-0170-9065,gname='Yong-Gang',sname='Zheng']{Yong-Gang Zheng}
\affiliation{Department of Physics, Yunnan Normal University, Kunming, Yunnan, 650092, People's Republic of China}
 \email[]{ynzyg@ynu.edu.cn}

\author[orcid=0000-0003-4773-4987,gname='Qingwen',sname='Wu']{Qingwen Wu}
\affiliation{Department of Astronomy, School of Physics, Huazhong University of Science and Technology, Wuhan, Hubei, 430074, People's Republic of China}
\email[]{qwwu@hust.edu.cn}

\begin{abstract}

We report the detection of a tentative, narrow spectral feature in the GeV gamma-ray spectrum of the blazar 4FGL J1754.2$+$3212, using approximately 16.5 years of observations from the Fermi Large Area Telescope. A detailed likelihood analysis of the time-averaged spectrum reveals a localized, line-like excess at an energy of $\sim$31 GeV, with a local statistical significance of $\sim$3$\sigma$ (3 degrees of freedom), without corrections for multiple trials and the look-elsewhere effect. This feature is not well described by standard smooth continuum models such as a single log-parabola. The addition of a Gaussian component significantly improves the spectral fit. If confirmed by future observations with higher sensitivity, this feature would represent a novel phenomenon in blazar astrophysics. We explore its potential physical origins, which could involve exotic processes such as dark matter annihilation or photon-axion-like-particle oscillations, or extreme astrophysical mechanisms within a structured jet. Regardless of its ultimate nature, this finding highlights the potential of deep spectral studies of blazars to reveal non-standard physics and challenges conventional one-zone emission models. Definitive verification awaits observations by next-generation instruments, particularly the Very Large Area $\gamma$-ray Space Telescope (VLAST).

\end{abstract}


\section{Introduction} \label{sec:Introduction}

Blazars, a prominent subclass of active galactic nuclei (AGNs), are characterized by relativistic jets aligned close to our line of sight \citep{1995PASP..107..803U}. Their strongly Doppler-boosted emission dominates the electromagnetic spectrum from radio to gamma-rays. The spectral energy distribution (SED) of blazars typically exhibits a double-humped, non-thermal structure. The low-energy component, spanning radio to X-rays, is universally attributed to synchrotron radiation from relativistic electrons. The origin of the high-energy component, peaking in the MeV to TeV regime, remains debated, with leading models divided into leptonic (e.g., inverse Compton scattering; \citealt{1992ApJ...397L...5M,1994ApJ...421..153S}) and hadronic (e.g., proton-synchrotron or proton-induced cascades; \citealt{2003APh....18..593M,2013ApJ...768...54B}) scenarios (see \citealt{2019Galax...7...20B,2022Galax..10...35P} for a review).

The $Fermi$ Gamma-ray Space Telescope, with its Large Area Telescope (LAT) sensitive from $\sim$20 MeV to over 300 GeV \citep{2009ApJ...697.1071A}, has revolutionized our understanding of the GeV sky. It has established that the spectra of most blazars are well described by simple power-law or smoothly curved (e.g., log-parabola) continua \citep{2015ApJ...810...14A,2020ApJ...892..105A,2020ApJS..247...33A,2023arXiv230712546B}. However, detailed analyses have occasionally revealed more complex features, offering critical diagnostics for jet physics. A seminal example is the detection of a significant spectral steepening (`break') at $\sim$2 GeV in the bright blazar 3C 454.3 \citep{2009ApJ...699..817A}. This phenomenon is commonly attributed to an intrinsic turnover in the particle energy distribution, absorption effects \citep{2009ApJ...699..817A}, or a superposition effect of energy spectra \citep{2021MNRAS.502.5875K}. Intriguingly, a converse spectral feature—a hardening or spectral turnover towards higher energies—has been identified in several nearby radio galaxies. Analyses of $Fermi$-LAT data for M 87 and Pictor A have provided evidence (at $\sim$2.5–5.8$\sigma$ significance) for a spectral inflection at several GeV, often interpreted as arising from two distinct emission components \citep{2019A&A...623A...2A,2025ApJ...988..268L}. Most remarkably, the flat-spectrum radio quasar S5 1027+74 was recently found to exhibit a prominent spectral break at $\sim$13 GeV, above which the $\nu F_\nu$ spectrum rises toward TeV energies—a feature never before observed in an FSRQ, interpreted as either the Klein-Nishina suppression of external Compton scattering off broad-line region photons or a $\gamma\gamma$ absorption signature imprinted by the broad-line region itself \citep{2025ApJ...991L...8P}. Such spectral hardening is extremely rare, with a population-level occurrence rate below 0.1\% \citep{2025A&A...703A.162D}, making the S5 1027+74 detection an exceptionally extreme case that manifests this phenomenon at energies an order of magnitude higher and in a quiescent state. Furthermore, a potential irregularity feature in the gamma-ray spectrum of the blazar B0516-621, interpreted as photon-axion-like-particle oscillations within the jet, has been reported \citep{2021JCAP...08..007Z}. Recently, \cite{2026arXiv260400579K} reported a persistent anomaly resembling a spectral line at approximately 1.5 GeV in three active galactic nuclei (AGNs), as well as a tentative double-excess signature observed in a single blazar object, which was discussed in \cite{2026arXiv260623488K}.
These discoveries underscore that complex spectral features in the GeV band serve as powerful probes of particle acceleration sites, emission mechanisms, and jet structures \citep[e.g.,][]{2012MPLA...2730030R,2019ARA&A..57..467B}.

Here, we report the discovery of a novel and tentative spectral feature of a different nature in the Fermi-LAT spectrum of the blazar 4FGL J1754.2$+$3212. Instead of a broad hardening or softening, our analysis reveals a localized, `line-like' excess at approximately 31 GeV. We present a detailed spectral analysis, assess its statistical significance, and discuss its potentially far-reaching implications for both astrophysical jet physics and fundamental physics.

\section{Data and Method} \label{sec:Data_Method}

\subsection{Data Analysis} \label{sec:Data_Analysis}

We analyze approximately 16.5 years (from 2008 August 4 to 2025 February 4, MJD 54682.65527778–60710.65527778) of Pass 8 (P8R3) Fermi-LAT data for 4FGL J1754.2$+$3212. The data reduction and analysis are performed using the Fermipy package (v1.2.0, \citealt{2017ICRC...35..824W}) and the standard Fermi Science Tools (v2.2.0). Events within a $15^{\circ}$ region of interest (ROI) centered on the source are selected in the energy range from 100 MeV to 1 TeV. We apply standard data quality cuts ((DATA\_QUAL$>$0)\&\&(LAT\_CONFIG==1)) and exclude photons with a zenith angle greater than $90^{\circ}$ to reduce contamination from Earth's limb. The instrument response functions P8R3\_SOURCE\_V3 are used throughout the analysis.

The background model includes all sources from the Fermi-LAT Fourth Source Catalog (4FGL-DR4, \citealt{2020ApJS..247...33A,2023arXiv230712546B}) within the ROI, as well as the Galactic diffuse emission model (gll\_iem\_v07.fits) and the isotropic component (iso\_P8R3\_SOURCE\_V3\_v1.txt). The spectral parameters of sources within $5^{\circ}$ of the target and the normalizations of the diffuse components are left free during the likelihood fit. SED data points with a test statistic (TS) exceeding 25 were obtained for further investigation; otherwise, flux upper limits at a 95\% confidence level were determined.

\subsection{Significance Analysis}\label{sec:Significance_analysis}

To investigate a potential localized feature, we perform a detailed spectral fit comparing two models: (1) a null model consisting of a single log-parabola:
\begin{equation}
\frac{{\rm d}N}{{\rm d}E} = K \left (\frac{E}{E_0}\right )^{-\alpha -
\beta\ln(E/E_0)},
\label{eq:logparabola}
\end{equation}
where $\beta$ is the curvature parameter, $\alpha$ is the spectral slope at the pivot energy $E_0$ (where the error on the differential flux is minimal), and $K$ is the normalization; and (2) an alternative model consisting of a log-parabola plus a Gaussian component\footnote{\url{https://fermi.gsfc.nasa.gov/ssc/data/analysis/scitools/source_models.html}} to fit the line-like excess:
\begin{equation}
N_{\rm signal}(E) = {N_0\frac{1}{\sigma_{\rm signal}\sqrt{2\pi}}}e^{-\frac{(E-E_{{\rm signal} })^2}{2\sigma_{\rm signal}^2}},
\end{equation}
where $N_0$ is the line amplitude, $E_{\rm signal}$ is the line center energy, and $\sigma_{\rm signal}$ is the line width (standard deviation).

Using the \textit{pyLikelihood} module\footnote{\url{https://fermi.gsfc.nasa.gov/ssc/data/analysis/scitools/python_usage_notes.html}}, we perform a likelihood analysis with the pure log-parabola model (null hypothesis) to obtain the maximum likelihood value $L_{\rm null}$. We then perform another likelihood analysis with the log-parabola plus Gaussian model (signal hypothesis) to obtain $L_{\rm signal}$. The models are XML files containing all sources of interest within the ROI, created by the LATSourceModel package\footnote{\url{https://github.com/physicsranger/make4FGLxml}}.

By comparing $L_{\rm signal}$ with $L_{\rm null}$, we calculate the test statistic $\mathrm{TS} = 2(L_{\rm signal} - L_{\rm null})$, which approximately follows a $\chi^2$ distribution. The local significance ($\sigma$) of the signal is derived from $\sigma = \sqrt{\mathrm{TS}/n}$, where $n=3$ is the number of additional degrees of freedom introduced by the Gaussian parameters ($N_0$, $E_{\rm signal}$, $\sigma_{\rm signal}$).

\begin{figure}[tp!]
    \centering
    \begin{overpic}[width=0.99\textwidth]{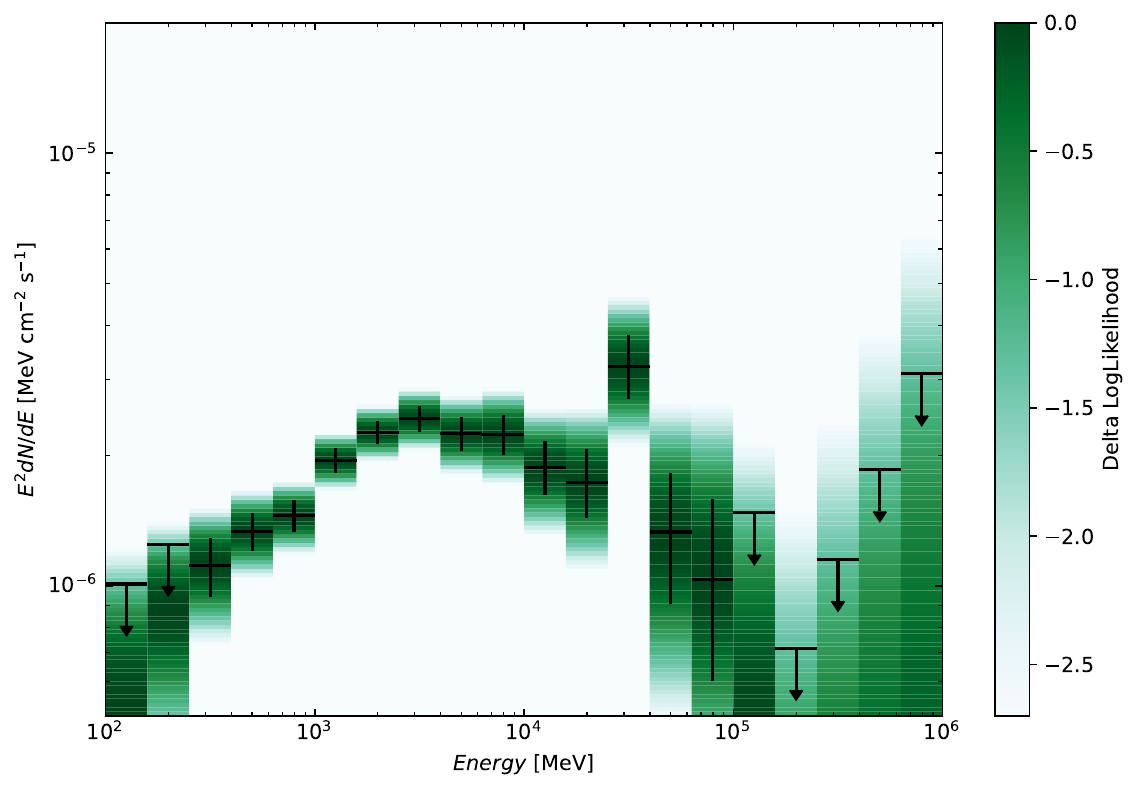}
    \put(10,43){\includegraphics[width=0.35\linewidth]{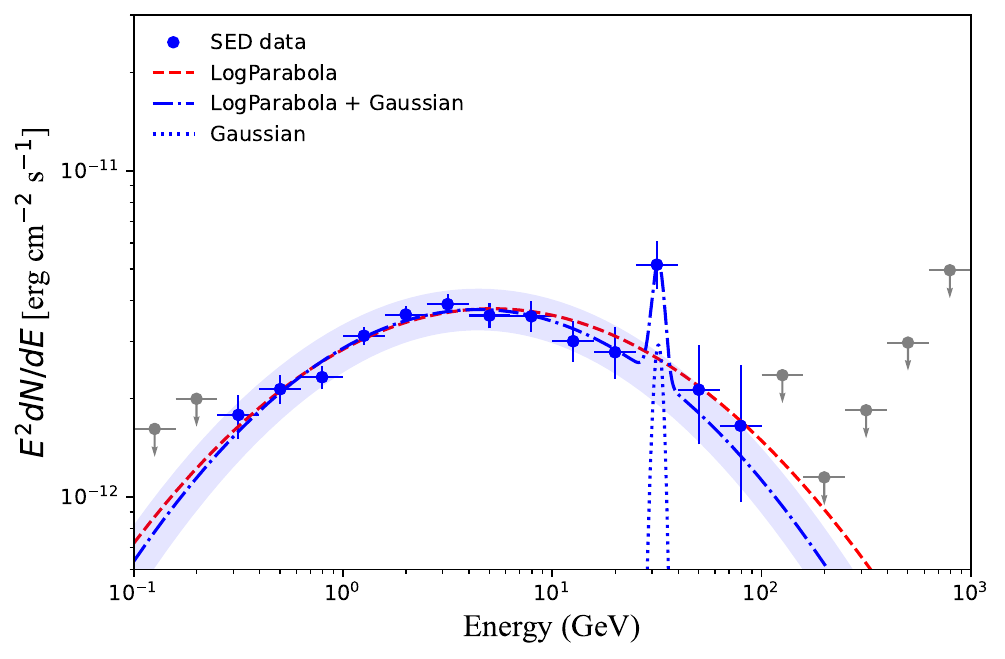}}
    \end{overpic}
    \caption{\textbf{Spectral energy distribution (SED) and line-like excess signal of 4FGL J1754.2$+$3212} in the 100 MeV to 1 TeV range, using 16.5 years of data from 2008 August 4 to 2025 February 4. The green color map illustrates the $\Delta$LogLikelihood for each SED point. In the inset panel, the SED is fitted with a pure log-parabola model (null hypothesis, red dashed line) and with a log-parabola plus Gaussian model (signal hypothesis, blue dot-dashed line). Blue points denote SED bins with TS $> 25$; gray downward arrows indicate upper limits for TS $< 25$. The blue dotted line represents the Gaussian excess component. The light-blue band shows the 95\% confidence interval for the continuum baseline.}
    \label{Fig_sed_01}
\end{figure}

\section{Results}

4FGL J1754.2+3212 that located at RA = 268.5532, Dec = 32.2007 is a BL Lac (BL Lacertae objects) type of blazar reported in 4FGL catalog \citep{2022ApJS..260...53A,2020ApJS..247...33A}.  It is associated with the source name RX J1754.1+3212, has not yielded a valid redshift obtained from spectroscopic observations reported in the literature. However, the redshift mean \citep{2016MNRAS.462.1775B} is estimated to be 1.09 based on the upper ($z_{\rm max} \simeq 1.61$) and lower ($z_{\rm min} \simeq 0.57$) limits of redshift reported by \cite{2013AJ....146..127S}. 

\subsection{Average SED and Spectral Excess}

Following standard binned maximum likelihood analysis procedures\footnote{\url{https://fermi.gsfc.nasa.gov/ssc/data/analysis/scitools/binned_likelihood_tutorial.html}}, we constructed the time-averaged SED using 20 logarithmically spaced energy bins. 
Figure~\ref{Fig_sed_01} presents the average SED of 4FGL J1754.2$+$3212 over 16.5 years. A prominent gamma-ray excess is evident around 31 GeV, superimposed on a smooth log-parabolic continuum. A single log-parabola (blue dot-dashed line) describes the continuum reasonably well but leaves clear positive residuals in the 25–40 GeV range, as seen in the inset panel. The alternative model (log-parabola plus Gaussian) provides a significantly better fit.

\subsection{Statistical Significance}

\begin{figure*}
\begin{centering}
\includegraphics[width=0.49\columnwidth]{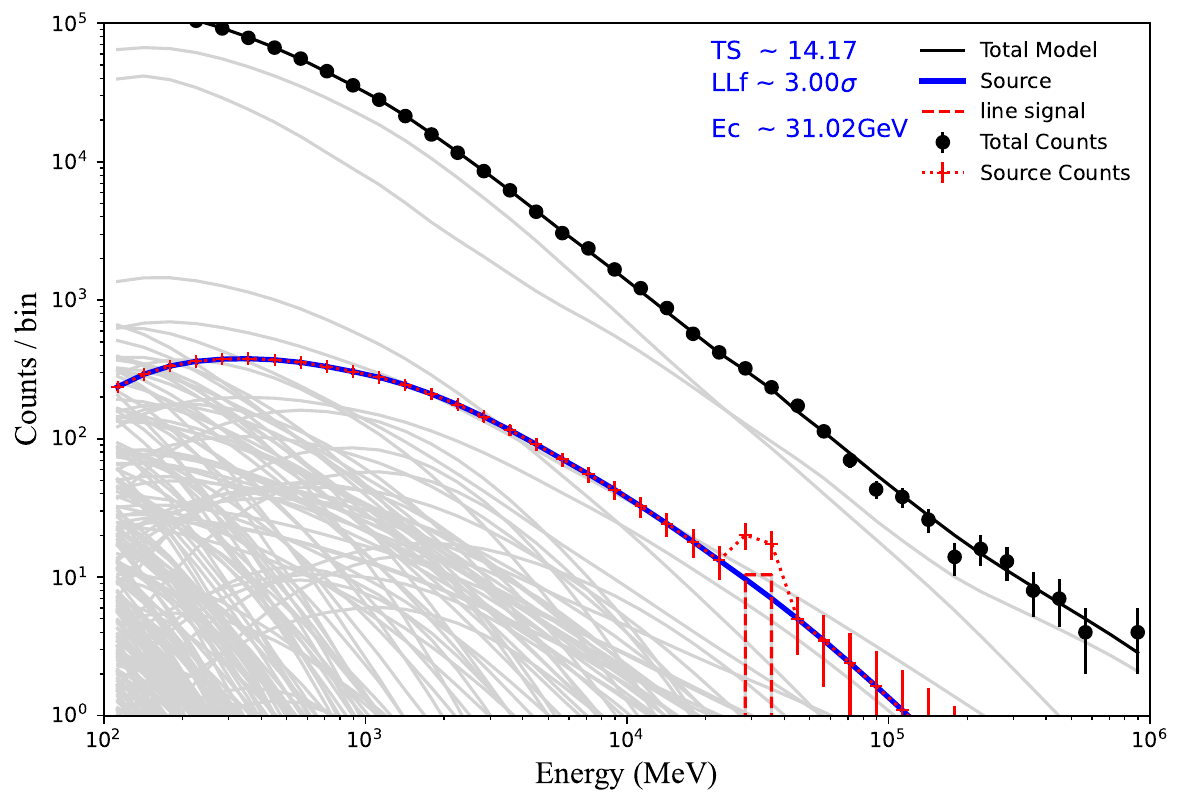}
\includegraphics[width=0.49\columnwidth]{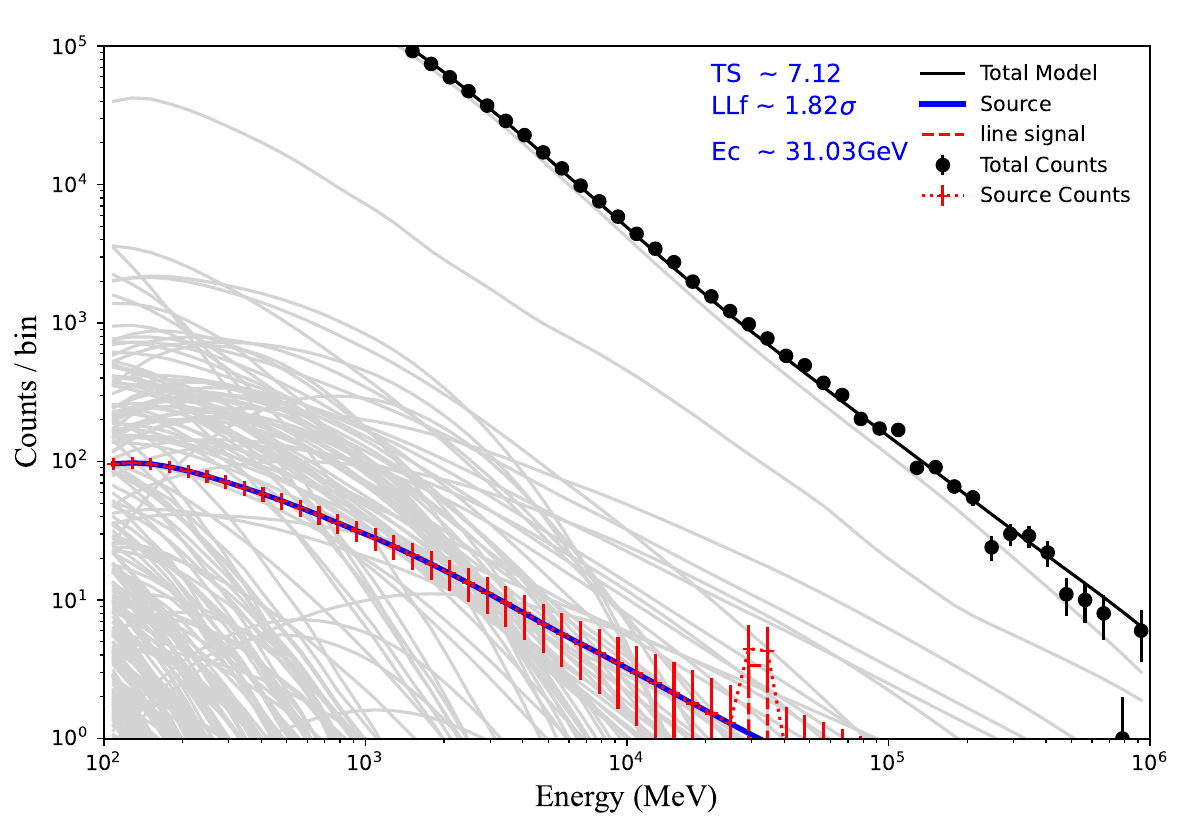}
\end{centering}
\caption{\label{figcount}
\textbf{Model-fitted Count spectra of  {4FGL J1754.2$+$3212} (left panel) and and  {4FGL J0057.9$+$6326} (right panel)}. It covers the energy range of 100 MeV to 1 TeV over a 16.5-year period - spanning from August 4, 2008, to February 4, 2025.
All three Gaussian parameters ($N_0$, $E_{\rm signal}$, $\sigma_{\rm signal}$) were kept free and scanned over a wide range (e.g., the full energy interval from 100 MeV to 1 TeV) in \textit{pylikelihood} analysis. Black dots and the solid black line show the total count spectrum and the total model from the \textit{pyLikelihood} analysis. Contributions of individual background sources in the model are drawn as light‑grey lines. Red dot‑dashed lines correspond to model‑fitted count spectra of the aimed sources. The blue solid line and the red dashed line represent the log-parabola model for the continuum and the Gaussian model for the line signal, respectively. The corresponding spectral‑fit parameters are listed in Table \ref{tab1}.}
\end{figure*}

To characterize the tentative spectral feature, we performed likelihood analyses with the pure log-parabola model and the log-parabola plus Gaussian model. During fitting, all three Gaussian parameters ($N_0$, $E_{\rm signal}$, $\sigma_{\rm signal}$) were free and scanned over a wide range (e.g., the entire energy spectrum spanning from 100 MeV to 1 TeV, see Figure \ref{figcount}). The best-fit Gaussian component is centered at $E_{\rm signal} = 31.02 \pm 0.13$ GeV with a width $\sigma_{\rm signal} = 1.6 \pm 0.4$ GeV (see Table~\ref{tab1}), basically consistent with the Fermi-LAT energy resolution at this energy. For the full 16.5-year dataset, we obtain $L_{\rm signal}= 1132860.84$ and $L_{\rm null}= 1132867.92$, yielding $\mathrm{TS} = 14.17$, which corresponds to a local significance of approximately $3\sigma$ (for three degrees of freedom,  without correction for trials and the look-elsewhere effect).

\begin{table*}[tp!]
\caption{\textbf{Spectral Fitting Parameters for the Line-like Excess Signal.}}
\begin{ruledtabular}
\begin{tabular}{ccccccccc}
 &&\multicolumn{2}{c}{Gaussian\footnote{Parameters from the Gaussian signal.}}
  &\multicolumn{3}{c}{Likelihood\footnote{Parameters related to the log-likelihood.}}\\
 \cline{3-4}\cline{5-8}
4FGL Name  &  Time interval (MET, s) &  $E_{\rm signal}$ (GeV) & $\sigma_{\rm signal}$ (GeV) & $L_{\rm null}$  & $L_{\rm signal}$ & $\mathrm{TS}$  & $\sigma$  & $\sigma_1$ \\  
\hline
                                 & 239557417 to 760376621 (\textbf{A}) & $31.02 \pm 0.13$ & $1.60 \pm 0.39$ & 1132867.92 & 1132860.84 & 14.17 & 3.00 & 3.76 \\
J1754.2$+$3212      & 239557417 to 760376621 (\textbf{B}) & $31.01 \pm 0.10$ & $1.44 \pm 0.23$ & 966109.65 & 966093.88 & 31.54 & 4.97 & 5.62 \\
                                 & 239557417 to 760376621 (\textbf{C}) & $31.03 \pm 0.12$ & $2.03 \pm 0.51$ & 811873.74 & 811856.48 & 34.53 & 5.25 & 5.87 \\
\hline
J0057.9$+$6326     & 239557417 to 760376621 (\textbf{A}) & $31.03 \pm 0.17$ & $1.63 \pm 0.87$ & -12573.55 & -12577.11 & 7.12 & 1.82 & 2.67 \\
\end{tabular}
\end{ruledtabular}
\par\noindent Note: Column 1: 4FGL source name. Column 2: Time interval (MET) covering 16.5 years from 2008 August 4 to 2025 February 4. (\textbf{A}): Full time interval. (\textbf{B}): Hypothetical case with twice the photon statistics, simulated by stacking the source data. (\textbf{C}): A selected partial time interval (see text). Columns 3 \& 4: Center energy and width of the Gaussian excess. Columns 5 \& 6: Log-likelihood values for the null and signal models. Columns 7 \& 8: Test statistic and corresponding local significance ($\sigma$) for three degrees of freedom. Column 9: Significance for one degree of freedom ($\sigma_1$).
\label{tab1}
\end{table*}

\section{Discussion} \label{sec:Discussion}

\subsection{Nature of the Spectral Feature}

The reported feature is a localized excess near 31 GeV, morphologically distinct from the broad spectral hardening seen in radio galaxies like M~87 and Pictor~A and blazar S5 1027+74 or the gradual softening break in 3C~454.3. Its sharp, line-like appearance (best-fit Gaussian width $\sigma_{\rm signal} \sim 1.6$ GeV,  compatible with the LAT's energy resolution) is the most intriguing aspect. The fitted width is basically consistent with the instrumental resolution, suggesting it may correspond to a monochromatic or intrinsically narrow emission process, though an intrinsically broad process cannot be ruled out. 
Moreover, it is important to note that the current limitations in SED data points prevent a precise determination of the spectral line center energy. This issue necessitates further investigation.

\begin{figure}[htp!]
\centering
\includegraphics[width=0.99\textwidth]{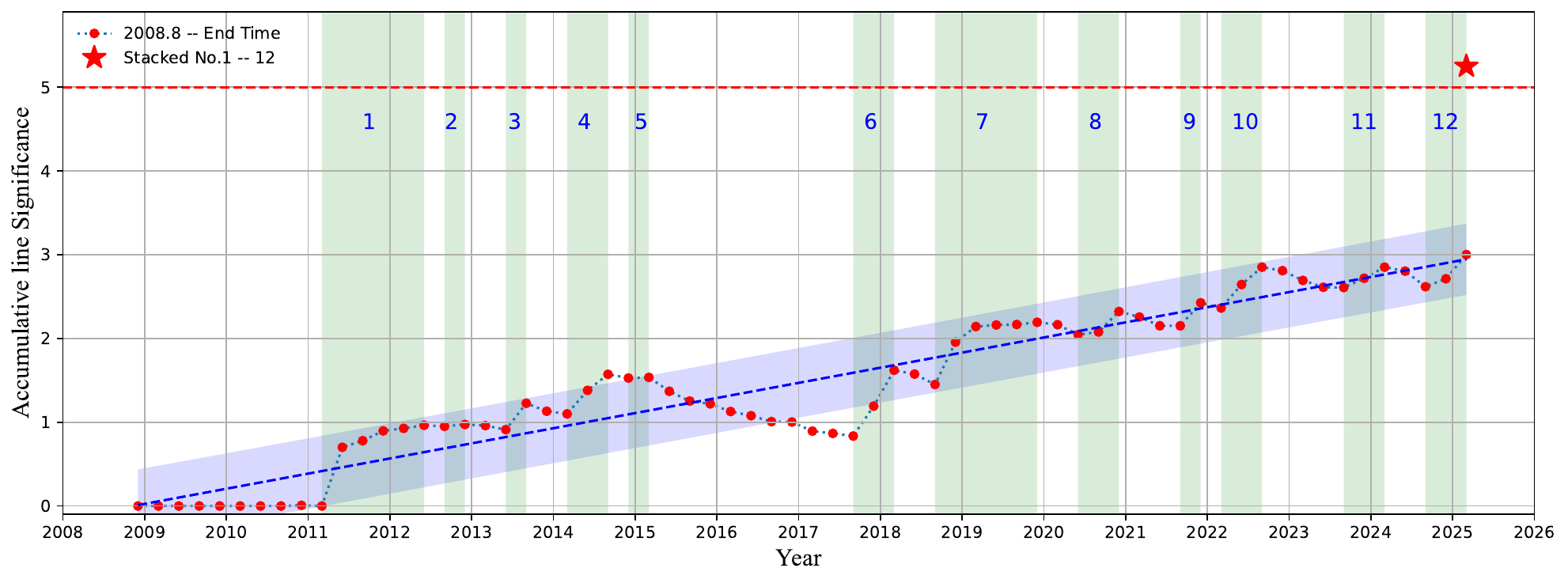}
\caption{\textbf{Local significance of the line-like excess in 4FGL J1754.2$+$3212 as a function of cumulative exposure.} Each point represents the significance computed for a time interval starting on 2008 August 4 and ending at the date shown, with durations accumulated in three-month steps. The blue dashed line shows a linear regression fit, with the gray-blue band indicating the 95\% confidence interval. The light-gray-green vertical bands highlight the selected time intervals used for subsequence \textbf{C} in Table~\ref{tab1}.}
\label{Fig_main_significance}
\end{figure}

The significance of the line-like signal depends on the photon accumulation time. Figure~\ref{Fig_main_significance} shows the local significance computed for different time intervals (each accumulated over three months). The cumulative significance generally increases with longer exposure. Extrapolating the trend suggests that with $\sim$30 years of data, the local significance may reach or exceed $5\sigma$. By hypothetically stacking the data to double the photon statistics, the significance reaches $\sim$5$\sigma$ (see case \textbf{B} in Table~\ref{tab1}).

Although the overall trend shows strengthening with accumulated photons, we observe that in some sub-periods the significance weakens with increasing exposure (similar to findings by \citealt{2021ApJ...920....1S,2024arXiv240711737F}), the cause of which requires further exploration. To improve the overall significance, we selectively filtered time intervals (similar to {{cherry-picking}}), removing periods where the signal weakened despite increased exposure and keeping those that enhanced it (light-gray-green bands in Fig.~\ref{Fig_main_significance}). Analyzing this selected subsequence yields $\mathrm{TS} = 34.53$, corresponding to a local significance of $\sim$5.25$\sigma$ (see case \textbf{C} in Table~\ref{tab1}). 
While this procedure bears similarity to cherry-picking, it yields, for the first time, a line-like GeV gamma-ray signal exceeding $5\sigma$ (\citealt{2024arXiv240601705C}).

Moreover, we have observed that this signal does not appear to be an isolated occurrence. In another {\it Fermi} source 4FGL J0057.9$+$6326  that is an unassociated source (see Figure \ref{figcount}), we detected a similar suspected signal with a central energy of approximately 31.03 $\pm$ 1.1 GeV, which aligns with the $\sim$31 GeV signal within the margin of error (see Table \ref{tab1}). This finding suggests that the signal may possess universal characteristics. However, it remains weak, exhibiting a maximum significance of  $\mathrm{TS} \simeq 7.1$ (corresponding to a local significance of about 1.8$\sigma$ for three degrees of freedom), thus, further research is necessary for its confirmation.

\subsection{Possible Physical Origins}

If confirmed as a narrow spectral line, the feature invites interpretations beyond standard astrophysics. The observed central energy is $\sim$31 GeV. In the source rest frame (co-moving with the jet), this energy scales as $E_{\rm co} \approx 31 \times (1+z) / \Gamma$ GeV, where $z$ is the redshift and $\Gamma$ is the jet's bulk Lorentz factor ($\Gamma=1$ if the emission does not originate in a relativistic jet). We consider several possibilities:

\subsubsection{Exotic Physics Probes}

\textbf{Electron-Positron Annihilation:} If the line originated from Doppler-boosted annihilation of electron-positron pairs (511 keV line in the co-moving frame), as suggested for GRB~221009A \citep{2024Sci...385..452R,2024SCPMA..6789511Z}, it would require an extremely high Lorentz factor of $\Gamma \sim 10^4$. This vastly exceeds typical values for Fermi blazars ($\Gamma \sim 20$; \citealt{2010A&A...512A..24S,2017Natur.552..374R}), making this interpretation highly unlikely.

\textbf{Proton-Antiproton Annihilation:} Direct annihilation of protons and antiprotons into two photons yields photons of $\sim$938 MeV each \citep{PhysRev_184_1415,1969PhRv..184.1415H,1967PhDT118H}. To boost this to $\sim$31 GeV in the observer frame for a source at redshift $z \sim 1.09$ \citep{2013AJ....146..127S,2016MNRAS.462.1775B} would require $\Gamma_{\rm boosted} \simeq 69$, again much larger than typical blazar Lorentz factors. This scenario is therefore also disfavored.

\textbf{Axion-Like Particles and New Jet Physics:} Photon-axion-like-particle oscillations in the intense magnetic fields of blazar jets could, under certain conditions, imprint oscillatory features onto the spectrum \citep{2016PhRvL.116p1101A,2021JCAP...08..007Z}, though no such oscillations are seen here. Alternatively, the line-like feature could stem from as-yet-unknown physical processes within the jet, posing a significant challenge to current radiation models and potentially revealing new jet physics.

\textbf{Dark Matter Annihilation or Decay:} AGNs are expected to harbor dense dark matter concentrations. Dark matter particles could be boosted to high velocities via scatterings with high-energy protons in relativistic jets \citep{2022PhRvL.128v1104W,2022JCAP...07..013G,2024JHEP...03..076X,2023arXiv230709460H,2024PhRvD.110a1701H,2025PhRvD.112e5004W,2025arXiv250907265B}. The signal could arise from boosted dark matter annihilation  or decay (e.g., $\chi\chi \to \gamma\gamma$ or $\chi\chi \to Z\gamma$). In the $\gamma\gamma$ channel, the dark matter particle mass would be $m_{\rm DM} \approx 31 \times (1+z)/\Gamma$ GeV. If the annihilation occurs within the jet with typical $\Gamma$, this would imply jets carrying boosted dark matter—a challenging scenario. If instead $\Gamma=1$, then $m_{\rm DM} \approx 31 \times (1+z)$ GeV. Assuming a value of $z \approx 0.38$, this would coincide with a reported $\sim$43 GeV line from galaxy clusters \citep{2016PhRvD..93j3525L,2021ApJ...920....1S,2024arXiv240711737F}, though this depends on the precise redshift. Decays of gravitino or glueball dark matter \citep{2002APh....16..451F,2014PhRvD..89k5017B,2014JCAP...10..023A,2016PhRvD..93k5025S} could also produce narrow line features.

\subsubsection{Astrophysical Continuum or Artifact?}

We must also consider less exotic possibilities. The $\sim3\sigma$ significance, while intriguing, is not conclusive. The feature could be a statistical fluctuation or a manifestation of an unmodeled, more complex continuum curvature (e.g., a second curved emission component). We have verified that our background modeling is robust and that no nearby flaring sources contaminated the ROI. Instrumental effects at this energy are expected to be smooth and unlikely to produce such a localized residual.

\subsection{Implications and Future Work}

The potential implications are profound. If linked to dark matter, 4FGL J1754.2$+$3212 could become a prime target for fundamental physics. If astrophysical, it points to highly specific and extreme conditions within blazar jets not captured by current one-zone models. Definitive confirmation or refutation is paramount and can be achieved through:
\begin{itemize}
    \item \textbf{Accumulated Fermi-LAT exposure:} Continued monitoring will reduce statistical uncertainties.
    \item \textbf{Next-generation gamma-ray observatories:} Missions like e-ASTROGAM \citep{2018JHEAp..19....1D}/AMEGO-X {\citep{2022JATIS...8d4003C}} (with superior MeV–GeV resolution) or the proposed VLAST telescope \citep{2022AcASn..63...27F} could critically examine such features.
    \item \textbf{Ground-based Cherenkov telescopes:} The Cherenkov Telescope Array (\url{https://www.cta-observatory.org/}, CTA) \citep{2013APh....43....3A}, with unprecedented sensitivity in a few tens of GeV to above 100 TeV range, is ideally suited for high-precision spectroscopy around 31 GeV.
    \item \textbf{Multi-wavelength/Multi-messenger campaigns:} Correlated variability studies and searches for associated neutrinos or ultra-high-energy gamma-rays could constrain hadronic models.
\end{itemize}

\subsection{Conclusion}

We have presented evidence for a tentative, line-like GeV excess at $\sim$31 GeV in the long-term Fermi-LAT spectrum of the blazar 4FGL J1754.2$+$3212, with a local significance of $\sim$3$\sigma$. This feature challenges description by standard smooth continuum models and adds a new, sharp morphological type to the growing diversity of complex GeV spectral features in AGN. While statistical and systematic uncertainties warrant caution, the potential physical interpretations—spanning from exotic particle physics to extreme astrophysical conditions in jets—are highly compelling. This discovery underscores the value of deep, spectrally resolved studies of AGN with current instruments and highlights a compelling target for future observations. We strongly encourage dedicated follow-up, in particular with the upcoming CTA and/or VLAST observatory, to unveil the true nature of this intriguing gamma-ray spectral feature.

\begin{acknowledgments}
We thank Z.-Q. Shen for assistance with data checking, and Y.-Z. Fan and J. Li for valuable discussions. This research has made use of data obtained through the High Energy Astrophysics Science Archive Research Center (HEASARC) Online Service (Fermi Science Support Center) provided by NASA Goddard Space Flight Center. This work is partially supported by the National Natural Science Foundation of China (Grant No.12163002, U1931203, U2031201, and 12363002) and the Liupanshui Science and Technology Development Project (Grant No. 52020-2024-PT-05) and the Discipline-Team of of Liupanshui Normal University (LPSSY2025XKTD07).
\end{acknowledgments}

\begin{contribution}

S.-J. Kang initiated and designed the study, and performed the data analysis together with Y. Yin. The manuscript was drafted by S.-J. Kang and revised by Y.-G. Zheng and Q. Wu. All authors discussed the results, contributed to the interpretation, approved the final version.

\end{contribution}

\facilities{Fermi (LAT).}
\software{Fermipy \citep{2017ICRC...35..824W}.}

\bibliography{sample701}{}
\bibliographystyle{aasjournalv7}

\end{document}